\documentstyle[12pt]{article}
\font\tenrm=cmr10

\def\fr#1#2{{{#1} \over {#2}}}

\def\ket#1{|{#1}\rangle}

\def\frac#1#2{{\textstyle{{#1}\over {#2}}}}

\def\lsim{\mathrel{\rlap{\lower4pt\hbox{\hskip1pt$\sim$}}
    \raise1pt\hbox{$<$}}}
\def\gsim{\mathrel{\rlap{\lower4pt\hbox{\hskip1pt$\sim$}}
    \raise1pt\hbox{$>$}}}
\def\sqr#1#2{{\vcenter{\vbox{\hrule height.#2pt
         \hbox{\vrule width.#2pt height#1pt \kern#1pt
         \vrule width.#2pt}
         \hrule height.#2pt}}}}

\newcommand{\beq}{\begin{equation}}
\newcommand{\eeq}{\end{equation}}
\newcommand{\bea}{\begin{eqnarray}}
\newcommand{\eea}{\end{eqnarray}}

\renewenvironment{thebibliography}[1]
 { \rm
   \begin{list}{\arabic{enumi}.}
    {\usecounter{enumi} \setlength{\parsep}{0pt}
     \setlength{\itemsep}{3pt} \settowidth{\labelwidth}{#1.}
     \sloppy
    }}{\end{list}}

\begin{document}
\titlepage

\title{Quantum-Mechanical Supersymmetry in Traps}

\author{V. Alan Kosteleck\'y and Neil Russell}

\begin{flushright}
{IUHET 369\\}
{August 1997\\}
\end{flushright}
\vglue 1cm
	    
\begin{center}
{{\bf QUANTUM-MECHANICAL SUPERSYMMETRY IN TRAPS%
\footnote{\tenrm
Presented by N.R.\ at the UIC Workshop on 
Integrable Models and Supersymmetry,
University of Illinois at Chicago,
June 1997}
\\}
\vglue 1.0cm
{V.\ Alan Kosteleck\'y and Neil Russell\\} 
\bigskip
{\it Physics Department, Indiana University\\}
\medskip
{\it Bloomington, IN 47405, U.S.A.\\}
\vglue 0.8cm
}
\vglue 0.3cm

\end{center}

{\rightskip=3pc\leftskip=3pc\noindent
We discuss the application of quantum-mechanical supersymmetry
to particle traps. 
The supersymmetric-partner wave functions may be used 
to describe a valence fermion in a trap system 
with an isotropic harmonic-oscillator potential.
Interactions with the core are incorporated analytically.
The close similarity of this approach to 
the application of supersymmetry in atomic systems 
is made explicit by means of 
a radial mapping between the two systems.

}

\vfill
\newpage

\baselineskip=20pt

Supersymmetry has been 
an active research area for well over two decades.
Despite this,
very few physical supersymmetries 
are experimentally known.
One is the
appearance of an effective radial potential 
in the context of atomic systems 
\cite{kn1,kn2}.
We discuss a possible further application 
in the context of particle traps.
More details are given in the two references 
listed under our names.

The isotropic harmonic oscillator has a radial equation 
admitting a supersymmetric partner.
The physical implications of this mathematical fact
may be investigated 
using trap systems for which 
an isotropic potential can be established.
In the Ioffe-Pritchard trap \cite{ip,prit} and 
the time-averaged orbiting-potential (TOP) trap 
\cite{paec},
this condition may be satisfied.
Both are neutral-particle traps
that use the interaction 
of the magnetic dipole moment 
$\vec{\mu}$ of 
the particle with a confining magnetic field.
The traps select all dipoles aligned 
opposite to the direction of the magnetic field
and draw them into the region of weakest field at the center.
The Ioffe-Pritchard trap is purely
magnetostatic, 
with the field provided by two coils
and four linear conductors.
The TOP trap comprises six coils,
four of which have alternating currents,
creating a high-frequency rotating magnetic field.
An averaging procedure
removes the time dependence,
yielding an effective magnetostatic potential.
With suitable choices of currents in 
the conductors of these systems \cite{bem,kr2},
it is possible to ensure isotropy 
of the potential energy
near the center of the trap, 
\begin{equation}
U(r) = \mu B_0 \left(1 + r^2/r_0^2 \right)
\label{eq1}
\enspace .
\end{equation}
Here, $r_0$ is a length characteristic of the trap system,
and $B_0 \neq 0$ is the magnitude of the magnetic field 
at the center of the trap.
The radial wave functions describing 
a single trapped dipole may be expressed in terms of
the generalized Laguerre polynomials $L^{(\alpha)}_{N}(z)$.
Ignoring a factor of $r$ 
that removes the first-order derivative 
in the differential equation, 
the functions are
\begin{equation}
W_{N,L}(r) = C_{N,L} \left(r/r_0\right)^{L+1} 
    \exp{\left(-r^2/2 r_0^2\right)}
     L_{N/2 - L/2}^{(L + 1/2)}\left(r^2/r_0^2\right)
\label{eq2}
\enspace .
\end{equation}
Here,
$L=0,\,1,\,2,\,\ldots$ is the angular momentum, 
and $N=L,\,L+2,\,L+4,\,\ldots$
is the principal quantum number.
Normalization is ensured via the constant $C_{N,L}$.
The full solutions of the harmonic oscillator
$\ket{N,L,M}$ behave as 
\begin{equation}
W_{N,L}(r) \, Y_{L,M}(\theta,\phi)
\label{eq2b}
\enspace ,
\end{equation}
where the azimuthal quantum number $M$ 
takes the usual values.

Details of supersymmetric quantum mechanics 
may be found elsewhere \cite{ni,wi,ak,cks}.
We use the term {\em bosonic sector}
to refer to the given radial system 
with  fixed angular momentum $L$
and spectrum shifted to have zero lowest-state energy.
The term {\em fermionic sector}
is used for the partner system.

One of our objectives is to 
regard the radial-equation fermionic sector 
as providing an effective potential
for an excited  valence particle in a trap.
It experiences not only the trapping potential,
but also interactions with a core of other particles.
Before motivating this application of supersymmetry further, 
let us consider the case of a valence
particle with angular momentum $L=0$
that is excluded from occupying levels 
below $N=2$
by a filled core.
This exclusion can only hold for fermions,
to which we restrict ourselves here.
If the interactions between the trapped dipoles
are small compared with the natural spacing of 
the energies in the oscillator,
the number of particles in the core 
may be found 
by counting the levels in 
the single-particle bosonic system
lying below that of the valence fermion.
There are four in this case. 
One is the ground state, $\ket{N=0,L=0,M=0}$,
and the other three can be labelled as
$\ket{N=1,L=1,M=0,\pm 1}$.
Similarly, the core would have $20$
fermions if the $L=0$ valence fermion was restricted 
to $N \geq 4$.

The physical interpretation of the fermionic sector
is motivated by several observations.
It is well known that the fermionic sector 
is degenerate with the bosonic sector
except for the lowest bosonic state,
which has no corresponding fermionic state.
One may visualize a situation in which 
{\em physically} such a corresponding state exists,
but is inaccessible to the valence fermion
because it is occupied by a core fermion.
The absence of a zero-energy state
in the fermionic sector of  
the {\em mathematical} formalism
reflects this physical picture.
The core need not be occupied by only one fermion,
since there might be others of different angular momenta.
Adopting this interpretation,
the Pauli principle is seen to underlie 
radial supersymmetry,
and it becomes natural to interpret
the fermionic sector as describing a valence fermion
in a multifermion system.
For $L=0$, 
the effective radial potential for the fermionic sector
differs from the corresponding one
for the bosonic sector by an expression 
that includes the term $\hbar^2 / m r^2$ 
where $m$ is the valence fermion mass.
This additional repulsion is thus consistent with
the meaning of the Pauli principle.

So, the effect of the supersymmetry is 
to fill an inner core with fermions.
This procedure must leave the angular momentum 
of the valence particle unchanged.
We therefore construct the full three-dimensional
wave functions from the product of 
the fermionic radial wave functions 
$W_{N_s-1,L+1}(r)$
and the same spherical harmonics as 
in the bosonic sector (\ref{eq2b}):
\begin{equation}
\mbox{W}_{N_s-1,L+1}(r)\, \mbox{Y}_{L,M}(\theta,\phi)
\label{eq3}
\enspace ,
\end{equation}
where 
$N_s=L+2,\,L+4,\,L+6,\,\ldots$
is the principal quantum number.
Even though the constant $L$ in the radial function
appears to have been shifted,
the angular momentum of the system is defined by 
the spherical harmonics and is unchanged.
These full wave functions 
(\ref{eq3})
for the fermionic sector 
differ from the full wave functions 
(\ref{eq2b})
for the three-dimensional 
isotropic  harmonic oscillator.

An alternative description for multifermion traps
could account for the filled core 
by using the standard harmonic-oscillator solutions 
(\ref{eq2b})
but requiring the values of the principal quantum number 
for the valence fermion 
to exclude numbers corresponding to the filled core.
However, 
the valence fermion would then be described by 
an incomplete set of states
and the lowest valence state 
would have too many nodes.
Both of these drawbacks are
absent for the
fermionic functions 
obtained via supersymmetry,
which form a complete
orthonormalizable set of states
and for which the lowest state 
has no nodes.
They therefore
resemble solutions for 
other conventional bound systems 
in quantum mechanics.

The number of fermions in the core of 
a particular trap described by 
the functions
(\ref{eq3})
depends on the angular momentum of the 
valence fermion.
For $L=0$, the principal quantum number 
takes values 
$N_s=2,\,4,\, 6,\,\ldots$ and 
by the reasoning considered above,
this trap has four core fermions.
The filling of another shell in the core
can be accomplished by
shifting the fermionic-sector spectrum
to have zero lowest-energy state,
thereby treating it as a new bosonic sector.
A new fermionic sector is then obtained
via the usual supersymmetry procedure.
This describes a trap with $20$ core fermions
and a valence fermion with zero angular momentum.
Further iterations of this procedure 
fills further shells, giving cores with
$56,\, 120,\, 220,\, \ldots$ fermions.
For valence fermions with $L=1$, 
the concept is the same,
and there are 
$1,10,35,84,\ldots$
core particles.
General formulae for these sequences 
can be obtained \cite{kr2}.
They assume only one spin orientation 
since the dipoles in the  Ioffe-Pritchard and TOP traps 
are oriented against the magnetic field.
The formulae differ for other trap systems.

Although it accounts for the Pauli principle, 
radial supersymmetry 
ignores interactions between 
the valence fermion and the core.
We discuss one method of incorporating 
interactions developed in analogy with
analytical supersymmetry-based quantum-defect theory 
for atomic systems \cite{kn}.
The modifications in the 
oscillator energy spectrum 
due to interactions plausibly generate the form 
\begin{equation}
E_{N^*} = \mu B_0 + \hbar \omega_0 \left(N^* + 3/2 \right)
\enspace ,
\label{eq4}
\end{equation}
for $\omega_0 =(2\mu B_0 / m r_0^2)^{1/2}$.
The modified eigenvalues are thus
incorporated via 
a shifted principal quantum number $N^*$. 
Denoting the shift by $\Delta = \Delta(N,L)$
and including also an integral shift $I = I(L)$,
we define $N^* = N+I-\Delta$.
Equivalently, with $N_s =N+2I$, 
we write $N^* = N_s-I-\Delta$.
If in addition we shift the angular momentum,
$L^* = L+I-\Delta$,
and add the effective potential 
\begin{equation}
V_{\mbox{\tiny EFF}}(r) = \fr{\hbar^2}{2m}
      \fr{L^*(L^* +1)-L(L+1)}{r^2}
    + \hbar \omega_0 (N-N^*)
\label{eq5}
\end{equation}
to the differential operator in the Schr\"odinger 
equation for (\ref{eq2}),
analytical radial wave functions
that correspond to the modified eigenspectrum
(\ref{eq4})
are obtained.
They may be expressed in terms of the functional form 
(\ref{eq2}) as $W_{N^*,L^*}(r)$.
This analytical defect theory 
extends the radial supersymmetry quite naturally, and
if the defects are switched off appropriately
the exact bosonic and noninteracting fermionic sectors 
are recovered.

The application of supersymmetry in traps
closely follows the application of supersymmetry in 
multi-electron atoms  
and in ions \cite{kn1,kn2}.
The similarity in the two applications 
is more than a mathematical parallel
and can be made explicit in the form of 
a mapping.
The existence of a natural correspondence 
between the radial three-dimensional
Coulomb problem and the radial harmonic oscillator 
in two or four dimensions
was first noted by Schr\"odinger more than fifty years ago
\cite{sch},
and since then has received much attention
\cite{bfr,cp,knt,kr}.
We consider Coulomb dimensions  $d > 1$, 
to avoid normalization issues associated with 
the one-dimensional case,
and oscillator dimensions $D\geq 1$.
For these arbitrary-dimensional cases, 
the radial equations may still be separated
\cite{lo}, 
and we write the radial solutions as
$W_{D,N,L}(r)$ for the oscillator
and $w_{d,n,l}(r)$ for the Coulomb case.
Details of these functions may be found elsewhere
\cite{kr}.
As an example, $W_{D=3,L,N}(r)$ is identical to (\ref{eq2}).
We adopt lower-case symbols for the Coulomb system
and upper-case symbols for the oscillator systems,
with an exception made for the oscillator radial variable $r$. 
In the Coulomb system, this convention gives 
angular momentum $l$ 
and principal quantum number $n$.

A natural mapping between these two radial systems 
exists subject to certain conditions on 
the dimensions, 
the angular momenta,
and the principal quantum numbers.
The relationship between the radial wave functions is
\cite{knt}
\begin{equation}
W_{D,N,L}(r) =  K_{d, n, \lambda} \, r^{-1/2}
     \ w_{d,n,l}\left((n+\gamma)r^2\right) 
\enspace ,
\label{eq6} 
\end{equation}
where $\gamma=(d-3)/2$ is 
a dimension parameter for the Coulomb system
that vanishes in the three-dimensional case.
The constant $K_{d,n,\lambda}$ is selected 
to preserve the normalization.
The restrictions on this correspondence may be expressed as
\begin{eqnarray}
D &=& 2d - 2 - 2\lambda \label{eq7}
\enspace , \\
N &=& 2n - 2 + \lambda \label{eq8}
\enspace , \\
L &=& 2l + \lambda  \label{eq9} 
\enspace ,
\end{eqnarray}
where it can be seen from (\ref{eq9}) that $\lambda$,
which gives an extra degree of freedom in the mapping,
has to be integer valued.
For $d=3$, $\lambda$ may equal zero or one,
yielding oscillator dimensions of $D=4,2$.
This is Schr\"odinger's original result.
Equation (\ref{eq7}) shows
there is no such correspondence between 
the physically interesting cases of 
the $D=3$ oscillator system 
and the $d=3$ Coulomb system.
The oscillator is limited to even
dimensions $D$ only.

To circumvent these dimensional restrictions on 
the correspondence between the exact systems, 
we broaden the class of systems considered
to include ones with analytical modifications
of the type introduced above 
for interactions between particles in a trap.
So, whereas there is no natural mapping between the single-particle
Coulomb and oscillator systems,
a map may exist
between a trap with {\em several} fermions
and the exact Coulomb system.

We allow for an integral shift $J$ 
in the oscillator dimension via
the definition and requirement $D^* = D+J \geq 1$.
If we also define a dimension parameter
$\Gamma^*=(D^* -3)/2$,
which vanishes for $D^*=3$,
then with the choice of effective potential
\begin{eqnarray}
V_{\mbox{\tiny EFF}} (r) &=& 
\fr{\hbar^2}{2m}
\fr{(L^* +\Gamma^*)(L^* +\Gamma^*+1)-(L+\Gamma)(L+\Gamma+1)}{r^2}
\nonumber \\
&& + \  \hbar \omega_0 (N - N^* + \Gamma - \Gamma^*)
\label{eq10}
\end{eqnarray}
we obtain a differential equation with 
analytical solutions $W_{D^*,N^*,L^*}(r)$.
The values of the parameters $\Delta, \ I,$
and $J$ are restricted 
if normalizability 
and orthogonality are desired
\cite{kr}.
With this broader class of oscillator radial systems,
the case of a mapping from the $D^*=3$ oscillator
to the $d=3$ Coulomb system becomes possible.
It is obtained by setting 
\begin{equation}
\Delta-I =\lambda-1/2
\enspace ,
\label{eq11}
\end{equation}
and since $\lambda$ can only take values zero or one,
the mapping requires 
a nonzero defect $\Delta$ in the oscillator.
Explicitly, the relationship 
and its constraints are
\begin{eqnarray}
 W_{3,N^*,L^*}(r) &=&
     K_{3,n,1/2} \,
     r^{-1/2} 
     w_{3,n,l} \left(n \, r^2\right) \enspace , \label{eq12}  \\
N^* &=& 2 n - 3/2     \enspace , \label{eq13}  \\
L^* &=& 2 l + 1/2     \enspace . \label{eq14}
\end{eqnarray}
Condition (\ref{eq13}) ensures that 
the entire stack of states of the one system
maps across to the entire stack for the other,
with the lowest states in the stacks identified with  
each other, 
the second lowest states with each other, 
and so on.

This mapping 
is not the only one possible
between the three-dimensional systems.
An alternative method 
involves allowing for an analytical modification
in the Coulomb system
instead of in the oscillator
and then following methods similar 
to those leading to
(\ref{eq12}).
Such an analytical quantum defect may be introduced 
by shifting the principal quantum number 
to give eigenenergies according to 
the well-known Rydberg formula $E_0/(n^*)^2$ 
\cite{rydb}, 
where $E_0$ is the ground state energy of the Coulomb system
and $n^* = n-\delta$, with $\delta = \delta(n,l)$.
Analytical solutions exist for this system, 
and their applications include 
the study of highly excited valence electrons
\cite{bk1,bk2,bk3,bkt1,bkt2}.
Since the Rydberg formula 
models the spectra of multi-electron atoms,
this option establishes a mapping 
between a single-particle trap
and a multiparticle atom 
such as an alkali-metal atom.
A more general third option
involves analytical defects in {\em both}
systems
and provides a mapping from a multiparticle trap
to a multi-electron atom. 
This most general form of the mapping
incorporates in special cases 
the bosonic sectors,
the fermionic sectors, 
and the quantum-defect sectors for various dimensions.
For example, 
in a mapping between the three-dimensional cases,
the generalization of 
(\ref{eq11}) can take the form
\begin{equation}
\Delta-I=2(\delta-i)+\lambda-1/2
\enspace ,
\label{eq15}
\end{equation}
where $i$ is the analogue for the  Coulomb problem of 
$I$ for the oscillator.

Unlike the defects $\Delta$ for traps, 
the defects $\delta$ have been measured for many atoms
\cite{hk}.
They depend on $l$
but are asymptotically independent of $n$ for large $n$.
This feature is attractive 
as it allows approximate orthogonality
of the solutions.
It would be equally attractive if it could be established 
for the oscillator system.
Indeed, if the oscillator defects $\Delta$ 
could be measured,
it would be amusing to know if
the constraint 
(\ref{eq15}) 
on the radial mappings between the 
three-dimensional systems
is consistent with the known atomic defects $\delta$.

\def\ijqc #1 #2 #3 {(19#2): Internat.\ J.\ Quantum\ Chem.\  
  {\bf #1}, #3 } 
\def\jmp #1 #2 #3 {(19#2): J.\ Math.\ Phys.\ {\bf #1}, #3 }
\def\jms #1 #2 #3 {(19#2): J.\ Mol.\ Spectr.\ {\bf #1}, #3 } 
\def\jpa #1 #2 #3 {(19#2): J.\ Phys.\ A {\bf #1}, #3 }  
\def\npb #1 #2 #3 {(19#2): Nucl.\ Phys.\ B{\bf #1}, #3 }
\def\prep #1 #2 #3 {(19#2): Phys.\ Rep. {\bf #1}, #3 } 
\def\pra #1 #2 #3 {(19#2): Phys.\ Rev.\ A {\bf #1}, #3 }
\def\prd #1 #2 #3 {(19#2): Phys.\ Rev.\ D {\bf #1}, #3 }
\def\prl #1 #2 #3 {(19#2): Phys.\ Rev.\ Lett.\ {\bf #1}, #3 }
\def\ibid #1 #2 #3 {(19#2): ibid., \rm {\bf #1}, #3 }
\def\pla #1 #2 #3 {(19#2): Phys.\ Lett.\ A {\bf #1}, #3 }

\end{document}